\documentclass[traditabstract]{aachanged}
\usepackage[english]{babel}
\usepackage{longtable}
\usepackage{graphicx}	
\usepackage{epsfig}
\usepackage{epsf}
\usepackage{amsmath}	
\usepackage{amssymb}	
\usepackage{xcolor,soul}
\usepackage{subcaption}
\usepackage{soul}
\usepackage[normalem]{ulem}

\newcommand{\CI}[0]{{}C\,{\sc i}\,}

\begin{document}

\title{\bf Influence of Radiative Pumping on the HD Rotational Level Populations
in Diffuse Molecular Clouds of the Interstellar Medium}
\titlerunning{Radiative Pumping on the HD}
\author{V.V. Klimenko$^{1}$\thanks{E-mail: slava.klimenko@gmail.com}, A.V. Ivanchik$^1$}
\authorrunning{Klimenko et al.}
\date{Received  March 14, 2020; revised March 14, 2020; accepted March 24, 2020}

\institute{\it{$^{1}$Ioffe Institute, Russian Academy of Sciences, \\ ul. Politekhnicheskaya 26,
St. Petersburg, 194021 Russia}}

\abstract{We present a theoretical calculation of the influence of ultraviolet radiative pumping on the excitation of the rotational levels of the ground vibrational state for HD molecules under conditions of the cold diffuse interstellar medium (ISM). Two main excitation mechanisms have been taken into account in our analysis: (i) collisions with atoms and molecules  and (ii) radiative pumping by the interstellar ultraviolet (UV) radiation field. 
The calculation of the radiative pumping rate coefficients $\Gamma_{\rm ij}$ corresponding to Drane's model of the field of interstellar UV radiation, taking into account the self-shielding of HD molecules, is performed.
We found that the population of the first HD rotational level ($J = 1$) is  determined mainly by radiative pumping rather than by collisions if the thermal gas pressure $p_{\rm th}\le10^4\left(\frac{I_{\rm{UV}}}{1}\right)\,\mbox{K\,cm}^{-3}$ and the column density of HD is lower than  $\log N({\rm{HD}})<15$.  
Under this constraint the populations of rotational levels of HD turns  out  to  be as well a more  sensitive  indicator of the UV radiation intensity than the fine-structure  levels of atomic carbon (\CI$^*$/\CI and \CI$^{**}$/\CI). We suggest that taking into account radiative pumping of HD rotational levels may be important for the problem of the cooling of primordial gas at high redshift: ultraviolet radiation from first stars can increase the rate of HD cooling of the primordial gas in the early Universe. 
}


\maketitle

\section{Introduction}

HD are the next most abundant\footnote{The abundance of CO is approximately the same as that of HD.} molecules in the Universe after H$_2$. Their number density in the cold phase of the neutral interstellar medium (ISM) of our Galaxy is lower than the number density of molecular hydrogen approximately by $5-6$ orders of magnitude (Snow et al. 2008). Since the HD lines corresponding to electronic-vibrational-rotational transitions fall into the ultraviolet (UV) wavelength range, they were first detected in our Galaxy only in observations on board the Copernicus space observatory (Spitzer et al. 1973, 1974) that had a UV telescope and then in observations with the FUSE orbital space telescope (Lacour et al. 2005). At present HD lines have been detected in 41 absorption spectra along the lines of sight toward nearby bright stars in our Galaxy (Snow et al. 2008). 

In observations of galaxies in the early Universe the UV HD lines fall into the optical range due to the cosmological redshift $z$, $\lambda^{\rm obs}=\lambda^{\rm em}(1+z)$, and can be detected by optical ground-based telescopes. HD absorption lines were first detected in the spectrum of the quasar Q\,1232$+$0815 in 2001 (Varshalovich et al. 2001). At present, about 20 HD absorption systems have been identified in damped Lyman-alpha (DLA) systems with $z>1.7$ in the spectra of quasars (see, e.g., Ivanchik et al. 2015; Kosenko and Balashev 2018).

The relative HD/H$_2$ abundance was investigated by Le Petit et al. (2002), Liszt (2015), Ivanchik et al. (2015), and Balashev and Kosenko (2020). These authors pointed out that the HD/H$_2$ ratio my be used as an indicator of the physical conditions in the cold ISM phase (the number density, ionization rate by of the cosmic-ray background, intensity of UV radiation, the dust content). An analysis of the populations of the HD rotational levels of the ground vibrational state can provide additional information about the physical conditions in the ISM. Like H$_2$, HD has a system of rotational-vibrational levels that are populated by collisions with atoms and molecules (mostly H, He, H$_2$, and e$^{-}$) and by radiative pumping (through the upper electronic levels). At the same time, there is a significant difference in relaxation dynamics between HD and H$_2$: owing to the higher symmetry of H$_2$, the lifetime of its excited states is greater than ones in HD by several orders of magnitude. Therefore, the molecular hydrogen transitions in the spectra of quasars are reliably detected for higher rotational levels, ${J=3-8}$ (Balashev et al. 2017), 
while transitions of HD from the first excited rotational level ${\rm J=1}$ were detected only in two cases, towards the quasars J\,0812$+$3208 at $z_{\rm abs}=2.626$ (Balashev et al. 2010) and J\,0843$+$0221 at $z_{\rm abs}=2.786$ (Balashev et al. 2017). Transitions from the upper levels ${J\ge2}$ for HD have not yet been observed. An analysis of the relative HD level population $N(J = 1)/N(J = 0)$ allowed the gas number density to be determined, $n = 240\,cm^{-3}$ in J\,0812$+$3208\,A (Balashev et al. 2010; Liszt 2015) and $n=260-380\,{\rm cm}^{-3}$  in J\,0843$+$0221 (Balashev et al. 2017). The authors neglected the effect of radiation pumping due to self-shielding of HD molecules from UV radiation, since these systems have a high HD column density.

In this paper we present the results of our calculation of the radiative pumping rate coefficients for the HD rotational levels. We determine the range of physical conditions in the ISM and HD column densities whereby radiative pumping contributes significantly to the excitation of HD rotational levels. Our calculation of the radiative pumping rate coefficients is described in Section\,2. In Section \,3 we analyze the effect of self-shielding. In Section 4 we compare the relative population of the HD level ${J=1}$ with other indicators of the physical conditions and present our analysis of the physical conditions in two molecular clouds towards J\,0812$+$3208 and J\,0843$+$0221. 

\section{Radiative pumping calculation}

The structure of the HD levels is similar to that of the H$_2$ ones, but there is also a significant difference: since the HD molecule has a dipole moment, the transitions between its levels with $\Delta J=\pm1$ are permitted and a larger number of levels turn out to be interconnected in the radiative cascade. Following the description of an H$_2$ radiative pumping calculation (Black and Dolgarno 1976), we calculated the radiative pumping of HD rotational levels. HD molecules absorb UV radiation and populate excited electronic states (${\rm B\,^1\Sigma_u}$ and ${\rm C\,^1\Pi_u}$), and then relax to the rotational-vibrational levels of the ground state ${\rm X\,^1\Sigma_g^+}$ (subsequently producing a rotational-vibrational cascade).

\subsection{Rotational-vibrational cascade for the levels of the ground electronic state}

The main parameters describing the distribution of level populations during relaxation to the ground electronic state are the cascade efficiency factors $a(\nu_0,J_0;J)$, which describe the probabilities to occupy a rotational level $J$ of the ground vibrational state $\nu = 0$ through a series of spontaneous transitions from an excited vibrational-rotational state $(\nu_0,J_0)$. The scheme described by Black and Dolgarno (1976) was used to calculate $a(\nu_0,J_0;J)$.

Suppose that some level $(\nu_0,J_0)$ is populated at a constant rate $Q(\nu_0,J_0)$ (${\rm cm^{-3}s^{-1}}$). Then, the equilibrium population of the level $(\nu_0,J_0)$ is defined as follows:
\begin{equation}
n(\nu_0,J_0) = Q(\nu_0,J_0)/A(\nu_0,J_0),
\end{equation}
where
\begin{equation}
A(\nu_0,J_0)=\sum\limits_{\nu''=0}^{\nu_0}\sum\limits_{J''=0}^{J_{\rm max}}A(\nu_0,J_0;\nu'',J'') \,\,\,[\mbox{s}^{-1}]
\end{equation} 
denotes the total probability of spontaneous transitions from the level $(\nu_0,J_0)$  to various levels of the ground electronic state. The probabilities of spontaneous dipole and quadrupole transitions $A(\nu_0,J_0;\nu'',J'')$ for the rotational-vibrational levels of the HD ground electronic state were calculated by Abgrall et al. (1982) for vibrational levels $\nu\le17$ and rotational levels $J\le J_{\rm max}=13$. In this paper we took into account the levels with  $J\le J_{\rm max}=13$ and $\nu\le13$. This is justified by the fact that in molecular clouds, under typical physical conditions, the populations of the overlying levels are negligible and their subsequent inclusion does not affect the radiative pumping rate coefficients. The equilibrium populations of the underlying levels $(\nu,J)<(\nu_0,J_0)$ are determined from the system of equations:
\begin{equation}
n(\nu,J)A(\nu,J)=\sum\limits_{\nu''=\nu}^{\nu_0}\sum\limits_{J''=0}^{J_{\rm max}} n(\nu'',J'')A(\nu'',J'';\nu,J).
\label{eq3}
\end{equation}
Assuming  $Q(\nu_0,J_0)=1\,cm^{-3}s^{-1}$ we can calculate the cascade efficiency factors $a(\nu_0,J_0;J)$
\begin{equation}
    a(\nu_0,J_0;J) =  \sum\limits_{\nu''=1}^{\nu_0}\sum\limits_{J''=0}^{J_{\rm max}} n(\nu'',J'')A(\nu'',J'';0,J).
\end{equation}
Since the population rate is constant, the normalization condition must be fulfilled: the number of molecules appearing per unit time at the excited level $(\nu_0,J_0)$ is equal to the number of molecules arriving at the levels of the ground vibrational state:
\begin{equation}
\sum\limits_{J=0}^{J_{\rm max}} a(\nu_0,J_0;J) = Q_0 = 1\,\,{\rm cm^{-3}s^{-1}.}
\end{equation}
The values of $a(\nu_0,J_0;J)$ were calculated for each pair $(\nu_0,J_0)$ of the ground electronic state ($\nu_0=1..13$, $J_0=0..J_{\rm max}$) and are given in Table\,\ref{cascade_factors} (for the first four vibrational levels).

\subsection{Radiative pumping rate coefficients $\Gamma(J_i,J_j)$}

To describe the fraction of the molecules at the ground vibrational level $\nu=0$ passed from a state ($\nu=0, J_i$) to a state ($\nu=0, J_j$) during radiative pumping, let us introduce the rate coefficients $\Gamma(J_i,J_j)$:
\begin{equation}
    \Gamma(J_i,J_j) = \sum\limits_{\nu_0=1}^{14}\sum\limits_{J_0=0}^{J_{\rm max}} Q^{J_i}(\nu_0,J_0)a(\nu_0,J_0;J_j) + Q^{J_i}(\nu=0,J_j),
    \label{eq5}
\end{equation}
where $Q^{J_i}(\nu_0,J_0)$ describes the excitation rate of the levels of the ground electronic state $(\nu_0,J_0)$ through spontaneous  \footnote{The rate of the induced transitions is much lower than the
rate of the spontaneous ones, and their contribution may be neglected.}  transitions from the levels of excited HD electronic states and is defined as follows:
\begin{equation}
Q^{J_i}(\nu_0,J_0) = \sum\limits_{B,C}\left[\sum\limits_{\nu'=0}^{40}\sum\limits_{J'=0}^{J'_{\rm max}} \frac{R(0,J_i;\nu',J')}{A^{tot}(\nu',J')}A(\nu',J';\nu_0,J_0)\right],
\label{eq6}
\end{equation}
where $\frac{R(0,J_i;\nu',J')}{A^{tot}(\nu',J')}=\frac{n(\nu',J')}{n(0,J_i)}$ are the relative equilibrium populations of the rotational-vibrational levels $(\nu',J')$ of states ($B$ and $C$) when excited from the level of the ground electronic state $(0,J_i)$, $R(0,J_i;\nu',J')$  is the excitation rate through the absorption of UV radiation, $A^{tot}(\nu',J')=A_c(\nu',J')+\sum\limits_{\nu_0,J_0}A(\nu',J';\nu_0,J_0)$ is the total probability of radiative transitions from level $(\nu',J')$ of the excited electronic states to the continuum and the rotational-vibrational levels  $(\nu_0,J_0)$ of the ground electronic state (see data in Abgrall and Roueff 2006).

Here and below, the superscript $'$  denotes the populations of the excited HD electronic states and the subscript $0$  denotes the levels of the ground electronic state.

The photoabsorption rate is defined by the following expression:
\begin{equation}
\begin{split}
&R(\nu'',J'';\nu',J') = \int_0^{\infty}\sigma_{ik}(\nu)c u_{\nu}(\nu)d\nu =\\
&=f_{ik}\frac{\sqrt{\pi}e^2}{mc}\int_0^{\infty} H(a,x) c u_{\nu}(\nu)d\nu \simeq f_{ik}\frac{\pi e^2}{m} u_{\nu}(\nu_{ik}),
\end{split}
\label{eq7}
\end{equation}
where $f_{ik}$ is the oscillator strength of the transition between states ($\nu'',J''$) ) and ($\nu',J'$),  $u_{\nu}(\nu_{ik})$ is the spectral UV radiation density inside the cloud at the transition wavelength $\left[\frac{\mbox{photons}}{\mbox{cm}^3 \mbox{Hz}}\right]$, $H(a,x)=\frac{a}{\pi}\int_{-\infty}^{+\infty}\frac{\exp(-y^2)}{(x-y)^2+a^2}dy$ is the Voigt function with parameters $a=\Delta \nu_R/\Delta \nu_D$ and $x=\frac{c}{b}\left(\frac{\nu -\nu_{ik}}{\nu_{ik}}\right)$. In the
optically thin case, the value of the integral is equal to the value of the function at the transition frequency. Thus, the photoabsorption rate is proportional to the UV photon density. The influence of shielding effect is considered in the next section.

To calculate the photoabsorption rate, we use the standard model of an interstellar UV radiation field (Draine 1978). In the wavelength range $<2000$\,\AA corresponding to the HD transition wavelengths, the UV radiation intensity by the number of photons 
$\left[\mbox{photons}/\mbox{s\,cm}^2\mbox{Hz\,sr}\right]$ is described by the following expression from Sternberg and Dalgarno (1995):
\begin{equation}
\begin{split}
&\phi(\nu)=I_{\rm UV}\times4\pi\times\left[8.49\times10^{-05}\left(\frac{\mbox{1\AA}}
{\lambda}\right) - \right.\\
&\left.-0.13666\left(\frac{\mbox{1\AA}}{\lambda}\right)^2 + 54.482\left(\frac{\mbox{1\AA}}{\lambda}\right)^3\right]
\end{split}
\label{Idraine}
\end{equation}
In the case of an isotropic radiation, the spectral density is related to the intensity as $u_{\nu}=4\pi {\it I}_{\nu}/c$; the total radiation density in the range (912-1108\,\AA) is then $6.9\times10^{-4}\mbox{\,cm}^{-3}$. We introduce the scale factor $I_{\rm UV}$ to take into account the stronger radiation fields, then $I_\nu=I_{\rm UV} I_{\nu}^{\rm Draine}(\nu)$. The scale factor $I_{\rm UV}$ appears linearly in Eqs. (\ref{eq5}-\ref{Idraine}) and, therefore, the radiative pumping rate coefficients $\Gamma(J_i,J_j)$ depend linearly on $I_{\rm UV}$. The values of $\Gamma(J_i,J_j)$ calculated for the
standard galactic background radiation ($I_{\rm UV}=1$) are given in Table\,\ref{UV_rates}.


\begin{table*}
\caption{Cascade efficiency factors $a(\nu_0,J_0;J)$ for the population of rotational levels $J$ of the HD ground vibrational state  $\nu = 0$. The data for the first four vibrational levels are given.} 
\label{cascade_factors}
\begin{tabular}{|c|c|c|c|c|c|c|c|c|c|c|c|c|}
\hline
\hline
$\nu_0$  & $J_0$ & \multicolumn{11}{c}{$\nu=0, J$}\\
& &  0 & 1 &2 &3 &4& 5& 6& 7&8&9&10 \\
\hline
1 & 0& 0.00000 &  0.98813 &  0.01187  & 0.00000  & 0.00000  & 0.00000 &  0.00000  & 0.00000 &   0.00000&   0.00000 &   0.00000 \\
1 & 1 &
   0.50542 &   0.00683  & 0.48193  & 0.00582 &  0.00000 &  0.00000 &  0.00000  & 0.00000  & 0.00000 &  0.00000 &  0.00000\\
1   &      2&
   0.00868  & 0.68212 &  0.00909  & 0.29625 &  0.00386 &   0.00000 &   0.00000 &   0.00000 &   0.00000 &   0.00000 &   0.00000\\
         1  &        3 &
   0.00031 &   0.02783 &   0.77772 &   0.01348 &   0.17807 &   0.00258 &   0.00000 &   0.00000 &   0.00000 &   0.00000 &   0.00000\\
         1  &        4 &
   0.00002 &   0.00205 &   0.06066 &   0.81772 &   0.01552 &   0.10233 &   0.00170 &   0.00000 &   0.00000 &   0.00000 &   0.00000\\
         1  &        5 &
   0.00000 &   0.00025 &   0.00732 &   0.10166 &   0.82001 &   0.01433  &  0.05534  &  0.00109 &   0.00000 &   0.00000  &  0.00000\\
         1  &        6 &
   0.00000 &   0.00004  &  0.00125  &  0.01738 &   0.14244 &   0.79907 &   0.01109 &   0.02804 &   0.00069 &   0.00000 &   0.00000\\
         1  &        7 &
   0.00000  &  0.00001  &  0.00028 &   0.00386 &   0.03168 &   0.17934 &   0.76388  &  0.00758 &   0.01295 &   0.00043  &  0.00000\\
         1  &        8 &
   0.00000 &   0.00000 &   0.00008 &   0.00104 &   0.00856 &   0.04847 &   0.20742 &   0.72433 &   0.00461 &   0.00521 &   0.00026\\
         1  &        9 &
   0.00000 &   0.00000  &  0.00002   & 0.00033 &   0.00270 &   0.01527  &  0.06537  & 0.22859 &   0.68353  &  0.00258   & 0.00161\\
         1  &       10 &
   0.00000 &   0.00000 &   0.00001 &   0.00012 &   0.00096 &   0.00543 &   0.02326  &  0.08133  &  0.24296 &   0.64456 &   0.00137\\
\hline
         2  &         0  &
   0.30069 &   0.40525  &   0.28838  &   0.00566  &   0.00003  &   0.00000  &   0.00000  &   0.00000  &   0.00000  &   0.00000  &   0.00000\\
         2  &         1  &
   0.17862  &   0.50603  &   0.22670  &   0.08687  &   0.00178  &   0.00001  &   0.00000  &   0.00000  &   0.00000  &   0.00000  &   0.00000\\
         2  &         2  &
   0.21897  &   0.23825  &   0.34929  &   0.16015  &   0.03262  &   0.00072  &   0.00000  &   0.00000  &   0.00000  &   0.00000  &   0.00000\\
         2  &         3  &
   0.00847  &   0.35223  &   0.26761  &   0.24523  &   0.11438  &   0.01178  &   0.00029  &   0.00000  &   0.00000  &   0.00000  &   0.00000\\
         2  &         4  &
   0.00043  &   0.02722  &   0.44704  &   0.28251  &   0.15811  &   0.08070  &   0.00387  &   0.00011  &   0.00000  &   0.00000  &   0.00000\\
         2  &         5  &
   0.00004  &   0.00261  &   0.05961  &   0.50253  &   0.28344  &   0.09471  &   0.05584  &   0.00117  &   0.00004  &   0.00000  &   0.00000\\
         2  &         6  &
   0.00000  &   0.00035  &   0.00896  &   0.10051  &   0.52455  &   0.27436  &   0.05283  &   0.03808  &   0.00033  &   0.00001  &   0.00000\\
         2  &         7  &
   0.00000  &   0.00006  &   0.00168  &   0.02092  &   0.14154  &   0.52238  &   0.25987  &   0.02778  &   0.02566  &   0.00010  &   0.00000\\
         2  &         8  &
   0.00000  &   0.00001  &   0.00039  &   0.00506  &   0.03766  &   0.17767  &   0.50509  &   0.24311  &   0.01379  &   0.01719  &   0.00004\\
         2  &         9  &
   0.00000  &   0.00000  &   0.00011  &   0.00142  &   0.01106  &   0.05710  &   0.20486  &   0.48024  &   0.22730   &  0.00659  &   0.01133\\
         2  &        10  &
   0.00000  &   0.00000  &   0.00003  &   0.00046  &   0.00365  &   0.01967  &   0.07695  &   0.22583  &   0.45512  &   0.21504  &   0.00324\\
\hline
            3  &         0  &
   0.31352  &   0.32714  &   0.32114  &   0.03735  &   0.00084  &   0.00001  &   0.00000  &   0.00000  &   0.00000  &   0.00000  &   0.00000\\
         3  &         1  &
   0.15907  &   0.52005  &   0.20309  &   0.11003  &   0.00760  &   0.00017  &   0.00000  &   0.00000  &   0.00000  &   0.00000  &   0.00000\\
         3  &         2  &
   0.19438  &   0.26521  &   0.37865  &   0.11198  &   0.04765  &   0.00208  &   0.00004  &   0.00000  &   0.00000  &   0.00000  &   0.00000\\
         3  &         3  &
   0.08035  &   0.30363  &   0.24546  &   0.28275  &   0.06696  &   0.02022  &   0.00061  &   0.00001  &   0.00000  &   0.00000  &   0.00000\\
         3  &         4  &
   0.00490  &   0.15290  &   0.37504  &   0.21850  &   0.19898  &   0.04143  &   0.00804  &   0.00020  &   0.00000  &   0.00000  &   0.00000\\
         3  &         5  &
   0.00033  &   0.01669  &   0.22275  &   0.41400  &   0.18316  &   0.13334  &   0.02669  &   0.00296  &   0.00007  &   0.00000  &   0.00000\\
         3  &         6  &
   0.00003  &   0.00195  &   0.03879  &   0.28083  &   0.42467  &   0.14925  &   0.08554  &   0.01793  &   0.00098  &   0.00003  &   0.00000\\
         3  &         7  &
   0.00000  &   0.00029  &   0.00682  &   0.06917  &   0.32097  &   0.41451  &   0.12186  &   0.05364  &   0.01244  &   0.00028  &   0.00001\\
         3  &         8  &
   0.00000  &   0.00005  &   0.00140  &   0.01643  &   0.10248  &   0.34411  &   0.39224  &   0.10140  &   0.03295  &   0.00888  &   0.00007\\
         3  &         9  &
   0.00000  &   0.00001  &   0.00034  &   0.00430  &   0.03050  &   0.13415  &   0.35216  &   0.36469  &   0.08745  &   0.02002  &   0.00638\\
         3  &        10  &
   0.00000  &   0.00000  &   0.00010  &   0.00129  &   0.00973  &   0.04814  &   0.16214  &   0.35466  &   0.34019  &   0.07683  &   0.00694\\
\hline
         4  &         0  &
   0.25095  &   0.40008  &   0.27775  &   0.06817  &   0.00300  &   0.00006  &   0.00000  &   0.00000  &   0.00000  &   0.00000  &   0.00000\\
          4  &        1  &
   0.21629  &   0.40000  &   0.28365  &   0.08385  &   0.01567  &   0.00053  &   0.00001  &   0.00000  &   0.00000  &   0.00000  &   0.00000\\
         4  &         2  &
   0.15787  &   0.34428  &   0.31837  &   0.12916  &   0.04621  &   0.00401  &   0.00011  &   0.00000  &   0.00000  &   0.00000  &   0.00000\\
         4  &         3  &
   0.11430  &   0.25709  &   0.30199  &   0.23724  &   0.06699  &   0.02128  &   0.00108  &   0.00003  &   0.00000  &   0.00000  &   0.00000\\
         4  &         4  &
   0.02680  &   0.20557  &   0.30149  &   0.25207  &   0.17001  &   0.03422  &   0.00954  &   0.00029  &   0.00001  &   0.00000  &   0.00000\\
         4  &         5  &
   0.00215  &   0.06094  &   0.28266  &   0.32037  &   0.19427  &   0.11776  &   0.01761  &   0.00415  &   0.00009  &   0.00000  &   0.00000\\
         4  &         6  &
   0.00018  &   0.00838  &   0.10774  &   0.35115  &   0.32582  &   0.14092  &   0.05492  &   0.00955  &   0.00132  &   0.00002  &   0.00000\\
         4  &         7  &
   0.00002  &   0.00111  &   0.02083  &   0.15014  &   0.36930  &   0.29852  &   0.10082  &   0.05310  &   0.00546  &   0.00069  &   0.00001\\
         4  &         8  &
   0.00000  &   0.00018  &   0.00409  &   0.04033  &   0.19056  &   0.37920  &   0.27401  &   0.07275  &   0.03526  &   0.00336  &   0.00027\\
         4  &        9  &
   0.00000  &   0.00004  &   0.00091  &   0.01047  &   0.06441  &   0.22284  &   0.37376  &   0.24868  &   0.05333  &   0.02343  &   0.00213\\
         4  &        10  &
   0.00000  &   0.00001  &   0.00024  &   0.00299  &   0.02096  &   0.09153  &   0.24894  &   0.36606  &   0.22721  &   0.03518  &   0.00687\\
\hline
\end{tabular}
\end{table*}

\subsection{Self-shielding}
\label{sect_shielding}
HD and H$_2$ molecules, along with atomic hydrogen H\,{\sc i}, in molecular clouds are known to absorb UV radiation in lines, thereby shielding the interior of the cloud from radiation at the frequencies of
the corresponding transitions (see, e.g., Draine and Bertoldi 1996; Wolcott-Green and Haiman 2011). In these works the self-shielding factor is calculated as the ratio of the total dissociation rates of molecules deep in the cloud and at the cloud boundary, $f_{\rm{shield}}(N_{\rm HD})=\xi_{\rm diss}(N_{\rm HD})/\xi_{\rm diss}(N_{\rm HD}=0)$\footnote{Here (and below) $N$ is the column density expressed in $cm^{-2}$.}. The dissociation of molecules occurs as a process accompanying radiative pumping, so that some of the excited molecules relax to the continuum (about 15\%) and are destroyed, while the other
ones (about 85\%) pass to the excited levels of the ground electronic state. Thus, as a result of self-shielding, the dissociation rate of molecules ($\xi_{\rm diss}$), along with the excitation rate of the ground state levels $Q^{J_i}(\nu_0,J_0)$, weaken in the same way. We used the expression for the shielding factor from Draine and Bertoldi (1996). It gives zero shielding at the cloud boundary, in contrast to the approximation proposed by Wolcott-Green and Haiman (2011):
\begin{equation}
\begin{split}
f_{\rm shield}(x,D)= &\frac{0.965}{\left(1+x/D\right)^2} + \frac{0.035}{\sqrt{1+x}}\times\\
&\times\exp(-8.5\times10^{-4}\sqrt{1+x}),
\end{split}
\end{equation}
where $x=N({\rm{HD}})/8.465\times10^{13}\mbox{\,cm}^{-2}$  is the normalized column density and $D=b/10^5\mbox{\,cm\,s}^{-1}$ is the Doppler parameter.   

In our model the molecular cloud is described by a plane-parallel slab irradiated by a uniform interstellar radiation background on both sides. We assume the radiation flux to be uniform and incident normally to the cloud surface. The flux density on each of the cloud sides is $F_{\nu}= 2\pi I_{\nu} I_{\rm{UV}}$. The cloud is divided into parallel layers, in each layer the UV radiation density $u_\nu(x)$ is calculated by taking into account the shielding of the radiation coming from both sides:
\begin{equation}
    u_{\nu}(x) =  \frac{2\pi I_{\nu} I_{\rm{UV}}}{c} \left(f_{\rm{shield}}[N_{\rm{HD}}(x)] + f_{\rm{shield}}[N_{\rm{HD}}(l_c-x)]\right),
\end{equation}
where $l_c$ is the cloud size, $x$ is the coordinate along the line of sight (normally to the layer), and  $N_{\rm HD}(x)$ is the column density of molecules on the line of sight between the cloud edge and the depth  $x$. In the absence of shielding, the radiation density $u_{\nu}(x)=4\pi I_{\nu} I_{\rm{UV}}/c$ is constant and does not depend on $x$. Thus, the radiative pumping rate coefficients  $\Gamma(J_i,J_j)$ for molecules deep the cloud decrease by a factor
\begin{equation}
f_{\rm sh}(x) = \frac{1}{2}\left(f_{\rm{shield}}[N_{\rm{HD}}(x)] + f_{\rm{shield}}[N_{\rm{HD}}(l_c-x)]\right)
\end{equation}
compared to the unshielded case.

\section{Excitation of HD levels}
\label{Excitation_calc}

In equilibrium the populations of the HD rotational levels in the ground vibrational state are described by a system of linear equations:
\begin{equation}
\begin{split}
    \sum\limits_{i\neq j} N_i&\left(\sum\limits_{q}n_{q} k^q_{ij} + A_{ij} + \Gamma_{ij}\right) =\\
    &=N_{j}\sum_{i\neq j}\left(\sum\limits_{q} n_{q} k^q_{ji} + A_{ji} + \Gamma_{ji}\right)
\end{split}
\end{equation}
where the indices $i$ and $j$ are the HD rotational level numbers, $q$ are the particles involved in the collisions (H\,{\sc i}, pH$_2$, oH$_2$, He, and electrons), $n_q$ are the particle number densities, and $k_{ij}^{q}$ are the collisional rate coefficients that are functions of the kinetic temperature. The collisional rate coefficients were taken from Flower et al. (2000) and Dickinson and Richards (1975). The particle number densities with respect to the total hydrogen number density $n_H^{\rm tot}=n_{\rm H} + n_{\rm H_2} + n_{\rm H^+}$ were assumed to be equal to their typical values measured in diffuse molecular clouds of our Galaxy: $n_{\rm He}/n_H^{\rm tot}=0.085$ (Asplund et al. 2009), the electron number density $n_e/n_H^{\rm tot}=10^{-4}$ (in molecular clouds, as a rule, it corresponds to the abundance of ionized carbon, ${\rm n_C/n_H^{\rm tot}\sim2\times10^{-4}}\times Z$, where $Z$ is the metallicity), $n_{H_2}/n_H^{\rm tot} = 0.2$  (a typical molecular fraction of the gas in clouds with a high column density of molecular hydrogen $\log N_{H_2}>19$ (see, e.g., Balashev et al. 2019), the ortho-to-para hydrogen ratio was assumed to be equal to the equilibrium one $9\times \exp(-E_{10}/kT_{\rm kin})$).

\begin{figure*}
\begin{center}
        \includegraphics[width=\textwidth]{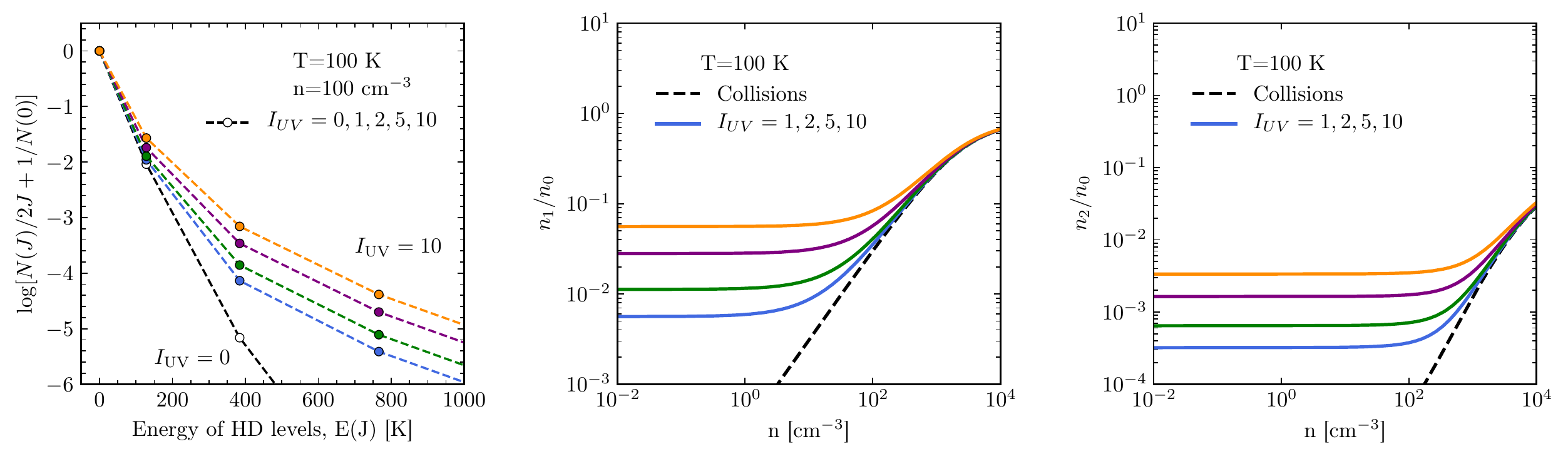}
        \caption{\rm Left panel represents populations of the HD rotational levels as a function of the energy of HD levels. The populations calculated for different intensities of the UV field $I_{\rm UV}$ = 0, 1, 2, 5 and 10, and coded in different colors: black, blue, green, violet, and orange. Right panel: the populations of HD rotational levels $J=1$ and $J=2$ relative to the ground level $n_J/n_0=n(J)/n(J=0)$ are shown as a function of the number density.}
        \label{HDpopratio}
\end{center}
\end{figure*}

The populations of the first and second HD rotational levels as functions of the number density, temperature, and UV background intensity are shown in Fig.\,\ref{HDpopratio}. We find that radiative pumping increases (by more than 10\%) the population of the first HD rotational level ${\rm J=1}$ in molecular clouds with a thermal gas pressure lower than
\begin{equation}
    p_{\rm{th}} = n T_{\rm kin} <10^4\left(\frac{I_{\rm UV}}{1}\right)\mbox{K\,cm}^{-3}.
    \label{eqPth}
\end{equation} 
This pressure is higher than the typical pressure in diffuse molecular clouds measured in the Milky-Way ($\log p=3.6\pm0.2$ by Jenkins and Tripp 2011), the Magellanic Clouds ($\log p=3.6-5.1$ by Welty et al. 2016), and galaxies at high redshifts observed in absorption as DLA systems in the spectra of quasars with $z=2-4$ 
($\log p_{\rm{th}}=4.0\pm0.5$ see, e.g., Balashev et al. 2019). At the same time, some systems exhibit higher UV intensity than the runoff field (e.g. Jenkins and Tripp 2011, Klimenko and Balashev 2020 and the references inside), which gives a higher threshold pressure. 

\subsection{Self-shielding effect}

\begin{figure*}
\begin{center}
        \includegraphics[width=\textwidth]{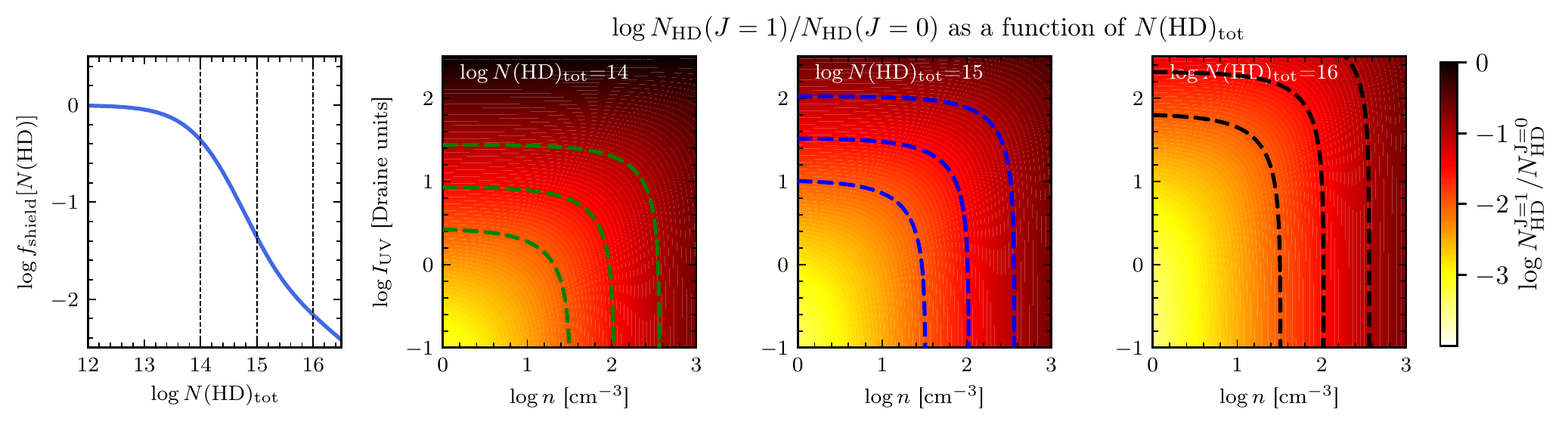}
        \caption{\rm 
        Left panel: the self-shielding function of HD. Right panels: the ratio $N_{\rm HD}(J=1)/N_{\rm HD}(J=0)$ as a function of the gas number density and intensity of UV radiation calculated for different total column density of HD cloud $\log N(HD_{\rm tot})=14, 15, 16$. Color indicates the logarithm of the ratio. The dashed lines indicate the isolines, corresponding to $N_{\rm{HD}}(J=1)/N_{\rm{HD}}(J=0)=1/100$, $1/30$, $1/10$.}
        \label{HDshield}
\end{center}
\end{figure*}

The self-shielding of HD molecules reduces the radiative pumping rate by the factor $f_{\rm sh}(x)$. We have calculated the ratio of the HD column densities at $J=1$ and $J=0$ rotational levels for various gas number densities and UV background intensities in clouds with $\log N_{\rm HD}=14, 15, 16$. The gas temperature was assumed to be 100\,K, corresponding to a typical kinetic temperature in diffuse molecular clouds (see, e.g., Balashev et al. 2019). The results are shown in Fig.\,\ref{HDshield}. The color gradient encodes the ratio $N_{\rm HD}(J=1)/N_{\rm HD}(J=0)$ as a function of the number density and UV intensity. Contours indicate the isolines corresponding to $1/100$, $1/30$, $1/10$. Thus, we conclude that at column densities $\log N_{\rm{HD}}<15$  pumping by UV radiation can contribute significantly to the HD rotational level populations even if the self-shielding of molecules is taken into account.

\begin{table*}
\caption{ Radiative pumping rate coefficients for HD rotational levels $\Gamma(J_i,J_j) \left[10^{-10}{\rm s^{-1}}\right]$ calculated for the standard interstellar UV radiation field in Draine model with $I_{\rm UV}=1$.}
\label{UV_rates}
\begin{center}
\begin{tabular}{|c|c|c|c|c|c|c|c|c|c|c|c|c|}
\hline
$J_i$  & \multicolumn{11}{c}{$J_j$}\\
&  0 & 1 & 2 & 3 & 4 & 5 & 6 & 7 & 8 & 9 & 10 \\
\hline
   0 &   0.839  &   1.323  &   1.109  &  0.358  &  0.077  &   0.010  &   0.001 &   0.000 &   0.000  &   0.000  &  0.000\\
   1  &  0.423  &   1.029  &   0.772  &   0.482 &   0.112  &  0.018  &  0.002  &  0.000  &  0.000  &  0.000  &  0.000\\
   2   & 0.374  &  0.787  &  0.915  &  0.481  &  0.289  &  0.047   & 0.006  &  0.000 &  0.000  &  0.000 &  0.000\\
   3  &  0.249  &  0.665  &  0.749  &  0.711  &  0.312   & 0.202  &  0.021  &  0.002  &  0.000  &  0.000 &  0.000\\
   4   & 0.160   & 0.468  &  0.711  &  0.652  &  0.609  &  0.220  &  0.148  &  0.011   & 0.001  &  0.000   & 0.000\\
   5   & 0.095  &   0.322  &  0.569  &  0.713  &  0.566  &  0.519  &  0.161  &  0.142  &  0.006  &  0.000  &  0.000\\
   6  &  0.049  &  0.202  &  0.433 &  0.613  &  0.664 &   0.478  &  0.468  &  0.117 &   0.133  &  0.003  &  0.000\\
   7  &  0.022  &  0.113  &  0.298  &  0.500  &  0.614  &  0.630  &  0.436  &  0.451  &  0.099  &  0.128  &  0.002\\
   8  &  0.008   & 0.054  &  0.179  &  0.374  &  0.557  &  0.631  &  0.611  &  0.397  &  0.428  &  0.087  &  0.117\\
   9   & 0.002  &  0.021  &  0.090  &  0.223  &  0.388  &  0.493 &  0.495 &   0.425 &   0.208 &   0.318 &   0.004\\
  10  &  0.001   & 0.008  &  0.038  &  0.114  &  0.230  &  0.326  &  0.342  &  0.289  &  0.256  &  0.095   & 0.164\\
\hline
\end{tabular}
\end{center}
\end{table*}

\section{Physical conditions in a molecular gas}

The population of the first HD rotational level is determined by collisions and radiative pumping, thus the ratio $N(J=1)/N(J=0)$ can be used to probe physical conditions (number density and intensity of UV radiation). These parameters can also be measured using the analysis of fine-structure levels of atoms and ions (Silva and Viegas 2002). The most useful indicator is neutral carbon \CI (for example, see study of \CI absorptions in the Milky-Way by Jenkins and Tripp 2011). 

\begin{figure*}
\begin{center}
        \includegraphics[width=\textwidth]{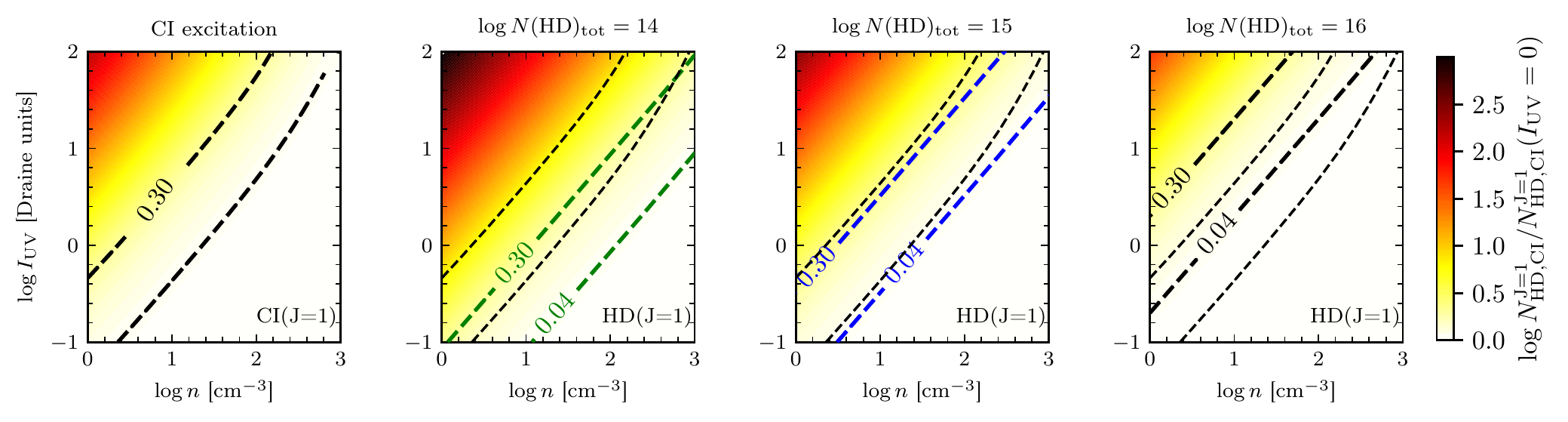}
        \caption{\rm 
        Calculation of the excitation of the C\,{\sc i}$^*$ fine-structure level (left panel) and HD $J=1$ level (right panels) performed for the range of UV intensity and number density and three column densities of HD: $\log N({\rm{HD}})=14, 15, 16$ (from left to right). The column density of HD is shown at the top of the panel. The color gradient indicates the logarithm of the ratio of (C\,{\sc i}$^*$ or HD $J = 1$) level populations excited by radiative pumping and collisions and only by collisions ($I_{\rm UV =0}$).
        The dashed and solid lines indicate the isolines corresponding to an increase of level populations by 10 and 100\% (or 0.04 and 0.3 dex). Black dashed lines correspond to \CI$^*$, green, blue and black thick dashed lines in right panels correspond to HD $J=1$.}
        \label{HDCIexc}
\end{center}
\end{figure*}

In Fig.\,\ref{HDCIexc} we compare sensitivity of the excitation of \CI fine-structure and HD rotational levels to rates of collisional and radiation pumping. We calculate the populations of \CI$^*$\footnote{Excitation of the second \CI$^{**}$ level shows approximately the same sensitivity as \CI$^*$} and HD (J=1) rotational levels for the range of physical conditions: $0< \log n <3$,  $-1<\log I_{\rm UV} <2$ and $T_{\rm kin}=100\,K$. 



It is usually believed that self-shielding strongly suppresses the radiative pumping of HD molecules. To check this assumption, we calculate the excitation of HD levels for three clouds with the total HD column density $\log N({\rm{HD}})=14, 15, 16$. Due to a typically low \CI column density, the self-shielding of \CI atoms is usually neglected. We find that at a column density $\log N({\rm HD})\le14$ the excitation of HD(${\rm J=1}$) level is  several times more sensitive to the UV intensity than the excitation of the \CI$^*$ (see left and second panels). 
As $N_{\rm HD}$ increases, the shielding suppresses the radiative pumping efficiency. At HD column densities $\log N({\rm{HD}})\sim15$  the radiative pumping of HD ($J=1$) and
C\,{\sc i$^*$} levels gives similar excitation. In case of higher column density  $\log N({\rm{HD}})>15$ the radiative pumping of the HD (${\rm J=1}$) is suppressed.
 
At the same time, as it was shown in Section\,\ref{Excitation_calc}, the upper HD rotational levels $J\ge 2$ have a higher sensitivity to the UV background intensity than $J=1$ and could be a good indicator of the UV radiation intensity. However, at present, the these transitions have not yet been detected in absorption in our Galaxy (see, e.g., Snow et al. 2008) and high-redshift galaxies (see, e.g., Ivanchik et al. 2015). Detection and analysis of high J rotational HD levels will measure the UV intensity in molecular clouds. This problem may become possible in nearest future, such as the Extremely Large Telescope (ELT) with the HIRES spectrograph (Oliva et al. 2018) or Spectrum-UV (Shustov et al. 2018) are put in operation.

\begin{table}
\begin{center}    
    \begin{tabular}{c|c|c}
     QSO & J\,0812$+$3208 & J\,0843$+$0221\\
     \hline
        HD ($J=0$) & $15.70^{+0.07}_{-0.07}$ & $17.34^{+0.13}_{-0.37}$ \\
        HD ($J=1$) & $13.77^{+0.15}_{-0.15}$ & $15.87^{+0.72}_{-0.49}$ \\
     \hline
    \end{tabular}
    \caption{
    Column densities of HD molecules at the ground and first rotational levels in the DLA systems towards the quasars J\,0812$+$3208 (Balashev et al. 2010) and J\,0843$+$0221  (Balashev et al. 2017).}
    \label{tab3}
\end{center}
\end{table}

\begin{figure*}
\begin{subfigure}{0.5\textwidth}
  \centering
  \includegraphics[width=0.95\textwidth]{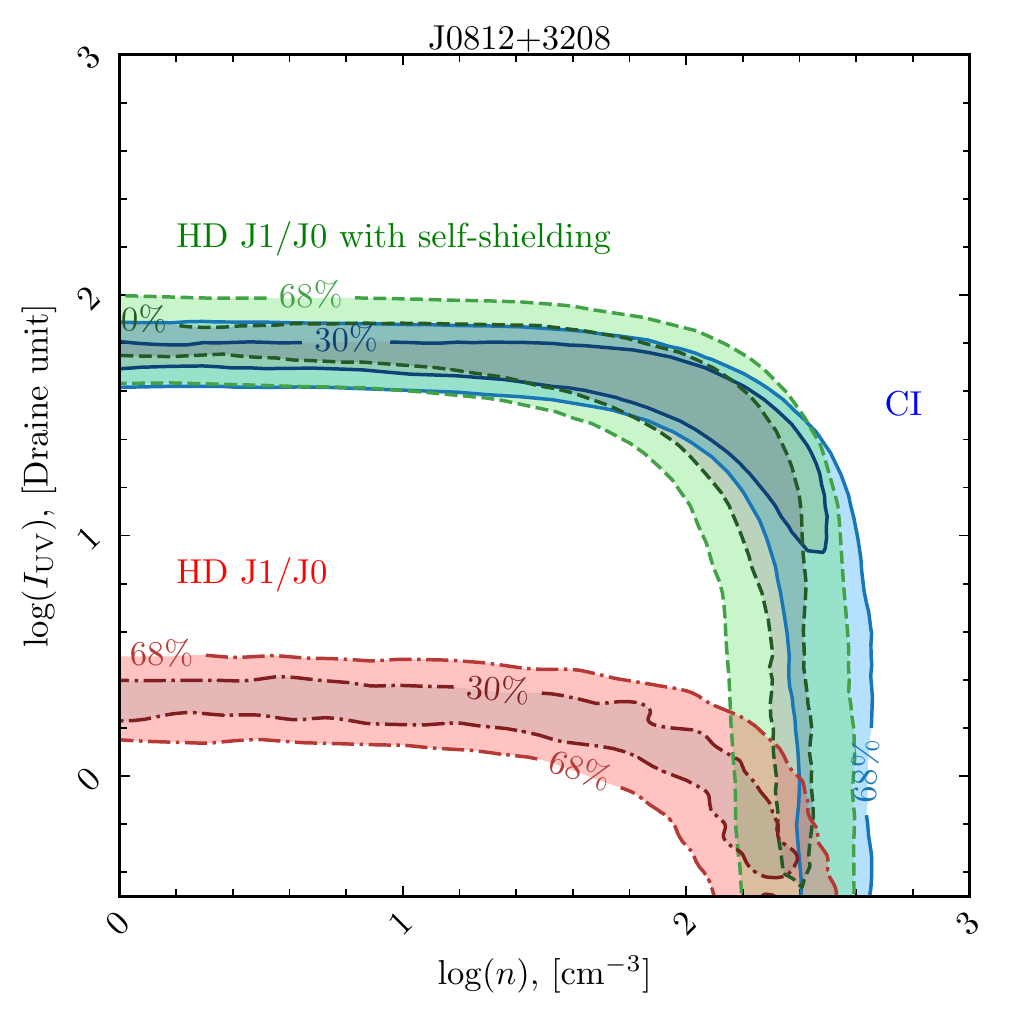}
\end{subfigure}%
\begin{subfigure}{0.5\textwidth}
  \centering
  \includegraphics[width=0.95\textwidth]{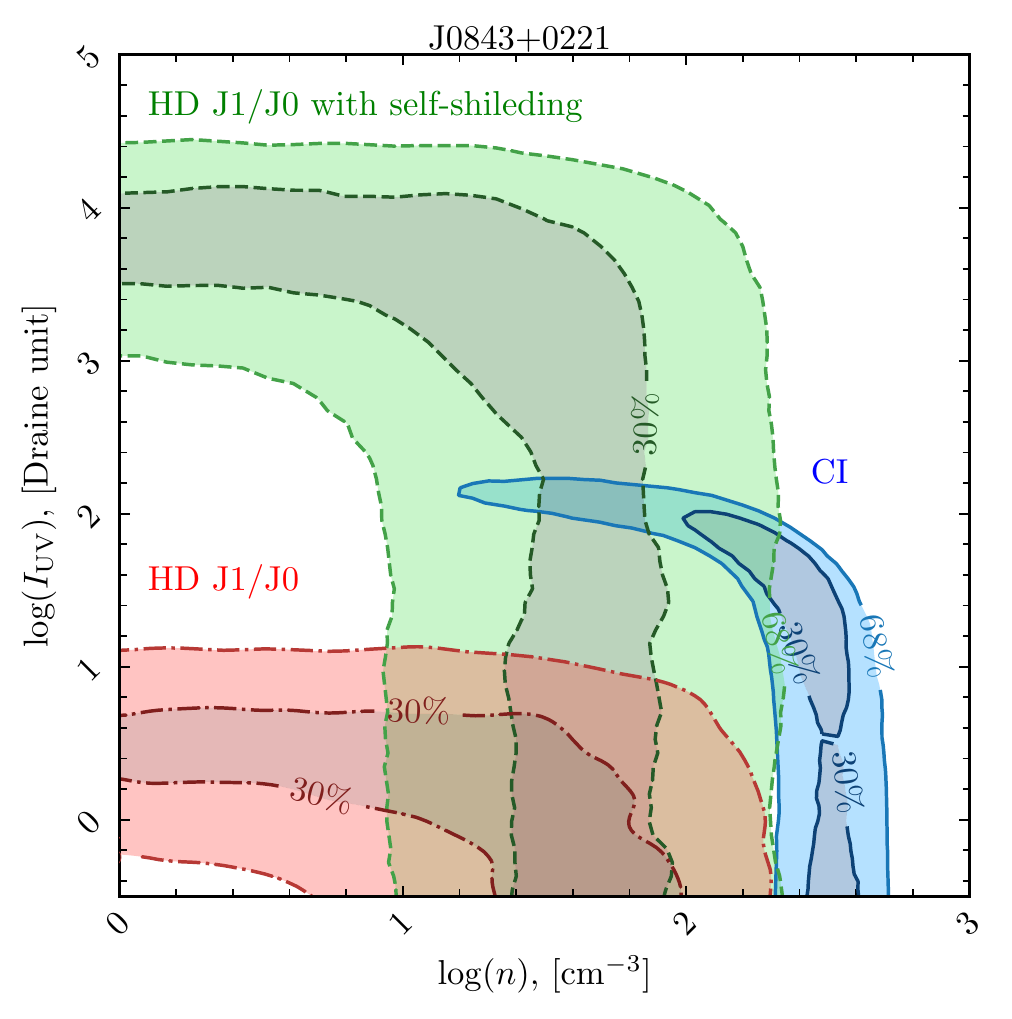}
\end{subfigure}
    \caption{
    The constraint on the gas number density and UV radiation intensity obtained by analysis of the relative population of the HD $J=1$ level (with and without self-shielding) and the fine-structure levels of C\,{\sc i} in the DLA system with $z=2.626$ in the spectrum of the quasar Q\,0812$+$3208 (a) and in the DLA system with $z=2.786$ in the spectrum of the quasar J\,0843$+$0221 (b). The contours indicate the probability density corresponding to 30\% and 68\% confidence levels. Even if the self-shielding effect is taken into account ($\log N({\rm{HD}})=15.7$  and 17.3 in Q\,0812$+$3208 and Q\,0843$+$0221, respectively), the constraint from the ratio $N(J=1)/N(J=0)$  for HD turns out to be comparable to the constraint obtained from the excitation of fine-structure levels of \CI.}
\label{hd_qso}
\end{figure*}

As an example, we constrain physical conditions using excitation of the HD rotational and \CI fine-structure levels in two DLA systems with high redshifts $z>2$ towards the quasars Q\,0812$+$3208A  (Balashev et al. 2010) and J\,0843$+$0221 (Balashev et al. 2017). In Fig.\,\ref{hd_qso} we present our estimates of the gas number density and UV intensity. The column density of HD rotational levels are given in Table\,\ref{tab3}. In case of J\,0812$+$3208\,A we find that HD and \CI give similar constraints. The kinetic temperature was equal to the excitation temperature H$_2$. If we assume that the UV intensity does not exceed 10 units of Draine field, we can estimate the number density, $n\sim240{\rm\,cm^{-3}}$. On the other hand, we can set an upper limit on the UV intensity $I_{\rm{UV}}<60$ units of Draine field. 
Therefore, even in case of high column density ($N_{\rm HD}=15.7$), radiative pumping produce a similar excitation of HD $J=1$ and \CI fine-structure levels.
We also show the constraint on the UV intensity with HD, when we neglect the self-shielding effect. The constrain on $I_{\rm UV}$ with HD could be about two orders of magnitude stronger than ones with \CI. In case of J\,0843$+$0221 constraints on $n_{\rm H}$ and $I_{\rm UV}$ obtained with HD are weaker than with \CI due to high uncertainty in the column densities of J=0 and J=1 HD levels. 

\section{Conclusions}

We consider the influence of radiative pumping by UV radiation on the excitation of lower rotational levels of the ground electronic state of HD molecules under physical conditions of the diffuse cold ISM. We calculate the rate coefficients for radiative pumping of HD molecules by background UV radiation in Draine model. The radiative pumping rate coefficients for the first 11 ($J=0-10$) rotational levels of the ground HD vibrational state are given in Table\,\ref{UV_rates}.

We show that at the edge of a molecular cloud, when the self-shielding of HD molecules may be neglected, the population of the first HD rotational level ($J=1$) is determined mainly by radiative pumping rather than by collisions if the thermal gas pressure satisfies the condition $p_{\rm th}\le10^4\left(\frac{I_{\rm{UV}}}{1}\right)\,\mbox{K\,cm}^{-3}$. Such conditions correspond to typical physical conditions of the cold phase of the diffuse ISM in local (Jenkins and Tripp 2011) and high redshift galaxies (Balashev et al. 2019, Klimenko \& Balashev 2020). The populations of the upper HD rotational levels with
$J\ge2$ are determined mainly by radiative pumping (the contribution of collisional pumping does not exceed 10\%). Measuring the populations of these levels allows to estimate the intensity of UV radiation in molecular clouds.
 
We consider the influence of self-shielding effect on the radiative pumping efficiency of the excitation of J=1 HD rotational level. We show that at $\log N({\rm{HD}}) < 15$ radiative pumping by a UV radiation of average interstellar intensity ($I_{\rm UV} = 1$ Draine unit) significantly excite the first HD level (${\rm J = 1}$), if the number density $n < 50\,{\rm cm^{-3}}$ and temperature $\sim100 \,{\rm K}$. The additional mechanism of excitation of HD molecules can be important in calculating the cooling of primordial plasma behind the shock fronts at the galaxy formation epoch. For example, the ionizing radiation of the first stars can increase populations of HD rotational levels, that increases the HD cooling rate.

We suggest that the population of HD rotational levels $N(J=1)/N(J=0)$ can be used to estimate the intensity of UV radiation and number density in diffuse molecular clouds of the ISM. At a column density $\log N({\rm{HD}}) < 15$ the ratio of column densities J1/J0 turns out to be more sensitive to the UV intensity and less sensitive to the number density than the the populations of \CI fine-structure levels. As an example, we estimated the physical conditions in two DLA systems at high redshifts towards the quasars Q\,0812$+$3208A ($\log N({\rm{HD}})=15.7)$ and Q\,0843$+$0221 ($\log N({\rm{HD}})=17.3$), in which the lines of HD transitions from ${\rm J = 1}$ were detected (Balashev et al. 2010, 2017). In case of  Q\,0812$+$3208A we determined the number density in the molecular cloud, $\sim240\mbox{\,cm}^{-3}$ and constrained the UV radiation intensity, $I_{\rm{UV}}<60$ Draine field units.

\section{ACKNOWLEDGMENTS}

This work was supported by the Russian Science Foundation (project no. 18-12-00301).

\newpage
\section*{References}

H. Abgrall,  E. Roueff, and Y. Viala, Astron. Astrophys. Suppl. Ser. {\bf 50},  505 (1982).\\
H. Abgrall and E. Roueff, Astron. Astrophys. {\bf445}, 361 (2006).\\
M. Asplund, N. Grevesse, A.J. Sauval, P. Scott, ARA\&A, {\bf 47}, 481 (2009).\\
S.A. Balashev, A.V. Ivanchik, D.A. Varshalovich, Astron. Lett. {\bf36}, 761 (2010).\\
S.A. Balashev and D.N. Kosenko, MNRAS {\bf492}, L45 (2020).\\
S.A. Balashev, V.V. Klimenko, P. Noterdaeme, J.-K. Krogager, D.A. Varshalovich, A.V. Ivanchik, P. Petitjean, R. Srianand et al., MNRAS {\bf490}, 2668 (2019).\\
S.A. Balashev, P. Noterdaeme, H. Rahmani, V.V. Klimenko, C. Ledoux, P. Petitjean, R. Srianand, A.V. Ivanchik et al., MNRAS {\bf470}, 2809 (2017).\\
J.H. Black and A.  Dolgarno, Astrophys. J. {\bf203}, 132 (1976). \\
Varshalovich D.A., Ivanchik A.V., Petitjean P., Srianand R.,  Ledoux C. Astron. Lett. {\bf27}, 683 (2001).\\
D.E. Welty, J.T. Lauroesch, T. Wong,  D.G. York, Astrophys. J. {\bf821}, 118 (2016).\\
A.M. Wolfe, E. Gawiser, and J.X. Prochaska, Astrophys. J. {\bf593}, 215 (2003). \\
J. Wolcott-Green and Z.  Haiman, MNRAS {\bf412}, 2603 (2011).\\
A.S. Dickinson and D. Richards, J. Phys. B: At. Mol. Phys. {\bf8}, 2846 (1975).\\
E.B. Jenkins and T.M. Tripp, Astrophys. J. {\bf734}, 32 (2011).\\
B.T. Draine,  Astrophys. J. Suppl. Ser. {\bf36}, 595 (1978).\\
B.T. Draine and F. Bertoldi, Astrophys. J. {\bf468}, 269 (1996).\\
A.V. Ivanchik, S.A. Balashev, D.A. Varshalovich, Klimenko V.V., Astron. Rep. {\bf59}, 100 (2015).\\
V. Klimenko, S.A. Balashev, A.V. Ivanchik, D.A. Varshalovich, Astron. Lett. {\bf42}, 137 (2016).\\
D.N. Kosenko and S.A. Balashev,  J. Phys. Conf. Ser. 012009 (2018), doi:10.1088/1742-6596/1135/1/012009.\\
S. Lacour, M.K. Andre, P. Sonnentrucker, F. Le Petit, D.E. Welty, J.-M. Desert, R. Ferlet, E. Roueff et al.,  Astron. Astrophys.  {\bf430}, 967 (2005).\\
F. Le Petit, E. Roueff, and J. Le Bourlot, Astron. Astrophys. {\bf390}, 369 (2002).\\
H.S. Liszt, Astrophys. J. {\bf799}, 11 (2015).\\
P. Noterdaeme, C. Ledoux, P. Petitjean, F. Le Petit, R. Srianand, and A. Smette, Astron. Astrophys. {\bf474}, 393 (2007).\\
P. Noterdaeme, R. Srianand, H. Rahmani, P. Petitjean, I. Paris, C. Ledoux, N. Gupta, S. Lopez, Astron. Astrophys. {\bf577}, 24 (2015).\\
E. Oliva, A. Tozzi, D. Ferruzzi,  M. Riva, M. Genoni, A. Marconi, R. Maiolino, L. Origlia, Proceed. SPIE {\bf10702}, 18 (2018).\\
T.P. Snow, T.L. Ross, J.D. Destree,  M.M. Drosback, A.G. Jensen, B.L. Rachford, P. Sonnentrucker, R. Ferlet, Astrophys. J. {\bf688}, 1124 (2008).\\
A. Sternberg and A. Dalgarno, Astrophys. J. Supp. Ser. {\bf99}, 565 (1995).\\
A.I. Silva and S.M.  Viegas, MNRAS {\bf329}, 135 (2002).\\
L. Spitzer, J.F. Drake, E.B. Jenkins, D.C. Morton, J.B. Rogerson, D.G. York, Astrophys. J. {\bf181}, L116 (1973).\\
L. Spitzer, W.D. Cochran, and A. Hirshfeld, Astrophys. J. Suppl. Ser. {\bf28}, 373 (1974).\\
D.R. Flower, J. Le Bourlot, G. Pineau des Forets,  E. Roueff, MNRAS {\bf314}, 753 (2000).\\ 
B. Shustov, A.I. Gomez de Castro, M. Sachkov,  J.C. Vallejo, P. Marcos-Arenal, E. Kanev, I.  Savanov, A. Shugarov, et al., Astrophys. Sp. Sci. {\bf363}, 62 (2018).\\





\end{document}